\documentclass[aps,prc,preprint,groupedaddress,superscriptaddress]{revtex4-1}
\usepackage[utf8]{inputenc}
\usepackage{graphicx}
\usepackage{amsmath, amssymb, bm, mathrsfs, longtable}
\begin{document}

\title{Nuclear Structure Features of Gamow-Teller Excitations}

\author{Vladimir Zelevinsky}
\affiliation{Department of Physics and Astronomy, Michigan State University, \\
East Lansing, Michigan 48824-2320, USA}
\affiliation{National Superconducting Cyclotron Laboratory, Michigan State 
University, East Lansing, Michigan 48824-1321, USA}

\author{Naftali Auerbach} 
\affiliation{Department of Physics and Astronomy, Michigan State University, \\
East Lansing, Michigan 48824-2320, USA}
\affiliation{School of Physics and Astronomy, Tel Aviv University, Tel Aviv 
69978, Israel}

\author{Bui Minh Loc} 
\affiliation{School of Physics and Astronomy, Tel Aviv University, Tel Aviv 
69978, Israel}

\date{\today}

\begin{abstract}

It is widely accepted that nuclear Gamow-Teller transitions are quenched; shell-model calculations
also showed a clear anticorrelation between the Gamow-Teller strength and the transition
rate of the collective quadrupole excitation from the ground state. We discuss the physics beyond this
observation. It is based on the existence of spin-orbit coupling that is responsible for the non-zero
probabilities of Gamow-Teller transitions in self-conjugate nuclei ($N=Z$). The shell-model calculations
in the $fp$-space demonstrate the effects of the gradual artificial removal of the spin-orbit coupling
that influences Gamow-Teller and quadrupole modes in opposite way. The realistic spin-orbit splitting moves 
the cumulative Gamow-Teller strength up and leads to stronger fragmentation; both trends are discussed 
in terms of simple symmetry arguments. Along with this process, the Gamow-Teller operator excites, 
in addition to the main line of $L=0$ states, states with $L=2$ which should be added, with 
the interference terms, to account for the total strength.
\end{abstract}

\maketitle

\section{Introduction}

Experimental and theoretical studies of weak interactions in general and nuclear Gamow-Teller (GT)
transitions specifically are in the focus of modern physics being important for nuclear structure
and reactions, astrophysics, particle physics and the search of phenomena outside the Standard Model.
In spite of many efforts, some basic problems related to nuclear GT transitions are still
controversial. Below we will try to address old questions on the crossroads of nuclear structure
and mechanisms of the GT dynamics in complex nuclei which still are not convincingly answered.

For a long time it  is claimed that the GT strength exciting the ground or
a low-lying nuclear state is significantly quenched compared to the standard estimates
\cite{blin67,wilkinson73,BM81,gaarde81}. The experimental studies typically find only about (60-70)\%
of the total strength. When such a reduction factor is introduced, the advanced shell-model calculations,
including the Monte Carlo studies, agree with what is observed, for example, in the $^{56}{\rm Ni}\,(p,n)$
charge-exchange reaction \cite{sasano11}. This subject was broadly discussed in the literature,
and, as stated in the old review article \cite{osterfeld92}, ``{\sl Both detailed nuclear structure
calculations and extensive analysis of the scattering data suggest that the nuclear configuration
mixing effect is the more important quenching mechanism, although subnuclear degrees of freedom
cannot be ruled out."}

One argument in favor of nuclear mechanisms behind the quenching is that the GT
strength considerably grows for the processes started in excited states $|\nu\rangle$. The shell-model
analysis of the GT strength for the $^{24}$Mg nucleus \cite{frazier97} shows a steady increase
of this strength as a function of excitation energy of the initial state. Apart from the statistical
effect of the level density, a considerable part of this increase comes from the suppression of
spatial symmetry and corresponding progress towards the Wigner ${\cal SU}(4)$ symmetry.

Another, qualitatively similar, phenomenon is the pronounced correlation, or rather anticorrelation \cite{AZZB93},
between the GT strength and the low-lying electric quadrupole (E2) strength. The same
conclusion follows from the graphs shown in a later work \cite{CJ15} on a different but related subject.
To the best of our knowledge, this effect is not sufficiently explained. This will be one of the subjects
of our discussion. We will find that the anticorrelation effect follows naturally as a consequence of
isospin invariance, fermionic antisymmetry of the wave functions, and spin-orbit coupling. Due to
spin-orbit splitting of single-particle levels, the total orbital momentum $L$ ceases to be an exact
quantum number so that the standard GT operator excites a superposition of $L=0$ and $L=2$ states.
It is  not clear if the usual experimental analysis correctly accounts for this fact which, however,
should be included in order to guarantee the total model-independent sum rule.

We will also confirm that the universal non-energy-weighted sum rule for the GT transitions is fulfilled
in the shell-model calculations only through many contributions of very weak transitions which can be
hardly visible in an experiment with  finite resolution and unavoidable background. The role of complicated
configurations in the saturation of the GT sum rule was stressed long ago \cite{klein85}. Below we show
exact results of the shell-model solution in the $fp$ space and add simple arguments based on the symmetry
considerations.

\section{Typical shell-model results}

We start with the results of typical shell-model calculations for few nuclei in the $fp$ shell. The normal
spin-orbit splitting in the FPD6pn shell-model version is 6.5 MeV between $f_{5/2}$ and $f_{7/2}$ levels and
$2$ MeV between $p_{3/2}$ and $p_{5/2}$ levels. The numerical experiment shown below, similarly to
Ref. \cite{AZZB93}, demonstrates the simultaneous calculation
of the total GT excitation probability $B^{-}$ from the ground state, and the quadrupole excitation rate
$B$(E2;$0^{+}\rightarrow 2^{+})$ for the lowest quadrupole collective excitation, as a function of the
gradually reduced spin-orbit splitting $\Delta\epsilon(f)=\epsilon(f_{5/2})-\epsilon(f_{7/2})$ to zero,
see Table 1.

We define the GT operators ${\bf V}^{\pm}$ as vectors with respect to spin variables, ${\bf s}=(1/2)\vec{\sigma}$,  carrying also  vector components in the nucleon isospin space ${\bf t}= (1/2)\vec{\tau}, \,\tau^{\pm}=\tau_{1}\pm i\tau_{2}$,
\begin{equation}
 {\bf V}^{-}=\,\frac{1}{2}\,\sum_{a}\vec{\sigma}_{a}\tau^{-}_{a},    \quad
 {\bf V}^{+}=({\bf V}^{-})^{\dagger}=\,\frac{1}{2}\,
\sum_{a}\vec{\sigma}_{a}\tau^{+}_{a},                                           \label{1}
\end{equation}
where the sums are taken over nucleons $a$; some useful algebraic definitions are included in Appendix A.

One can speak of the total GT strength of a given nuclear state $|\nu\rangle$ in the mother nucleus summed over all final daughter states,
\begin{equation}
B^{+}(\nu)=\,\frac{1}{2}\,\langle \nu|({\bf V}^{-}\cdot{\bf V}^{+})|\nu\rangle, \quad
B^{-}(\nu)=\,\frac{1}{2}\,\langle \nu|({\bf V}^{+}\cdot{\bf V}^{-})|\nu\rangle. \label{2}
\end{equation}
This definition, where the scalar product refers to the spin vectors, leads to the standard universal Ikeda sum rule, independent of the starting state $|\nu\rangle$,
\begin{equation}
B^{-}(\nu)-B^{+}(\nu)=\sum_{a}(\vec{\sigma}_{a})^{2}(\tau^{3})_{a}=3(N-Z).   \label{3}
\end{equation}
Here $|\nu\rangle$ is an arbitrary nuclear state below meson production threshold.
In particular, for nuclei with filled proton shells, such as $^{42-48}$Ca, the 
$B^{+}$ part is quite low, and the sum rule should be fulfilled mainly due to 
the $B^{-}$ part.

\begin{table}
\caption{The evolution of the total GT excitation probability from the ground 
state and the quadrupole excitation rate $B$(E2;$0^{+}\rightarrow 2^{+})$ from 
the ground state to the lowest quadrupole collective excitation
in $^{44}$Ti and $^{46}$Ti, in the process of gradual changing the spin-orbit 
splitting $\Delta\epsilon(\ell)= \epsilon(j=\ell-1/2)-\epsilon(j=\ell+1/2)$ 
from its realistic value to zero.
\label{Table1}}
\centering
\vspace{0.5cm}
\begin{tabular}{|r|rrrrrr|rrr|rrrr|} 
\hline
  & & & & & & & & $^{44}$Ti & & & $^{46}$Ti & & \\
  & $\epsilon(2p_{3/2})$ & $\epsilon(2p_{1/2})$ & $\Delta\epsilon(p)$ 
  & $\epsilon(1f_{7/2})$ & $\epsilon(1f_{5/2})$ & $\Delta\epsilon(f)$ 
  & $B^-$        & $B(\rm E2)$          & $ E(2^+)$
  & $B^-$      & $B^+$      & $B(\rm E2)$ 
  & $E(2^+)$ \\
\hline
0 & -6.495 & -4.478 & 2.017 & -8.388 & -1.897 & 6.491 & 1.26 & 699 & 1.30 & 
6.90 & 0.93 & 781 & 0.98 \\
1 & -6.495 & -4.478 & 2.017 & -8.088 & -2.197 & 5.891 & 1.05 & 734 & 1.27 & 
6.73 & 0.76 & 835 & 0.96 \\
2 & -6.495 & -4.478 & 2.017 & -7.788 & -2.497 & 5.291 & 0.87 & 764 & 1.22 & 
6.61 & 0.64 & 879 & 0.92 \\
3 & -6.495 & -4.478 & 2.017 & -7.488 & -2.797 & 4.691 & 0.71 & 793 & 1.17 & 
6.50 & 0.52 & 924 & 0.88 \\
4 & -6.495 & -4.478 & 2.017 & -7.188 & -3.097 & 4.091 & 0.56 & 820 & 1.14 & 
6.40 & 0.42 & 967 & 0.85 \\
5 & -6.495 & -4.478 & 2.017 & -6.888 & -3.397 & 3.491 & 0.43 & 843 & 1.11 & 
6.31 & 0.32 & 1008 & 0.82 \\
6 & -6.495 & -4.478 & 2.017 & -6.588 & -3.697 & 2.891 & 0.32 & 864 & 1.09 & 
6.23 & 0.25 & 1045 & 0.79 \\
7 & -6.495 & -4.478 & 2.017 & -6.288 & -3.997 & 2.291 & 0.23 & 880 & 1.07 & 
6.17 & 0.19 & 1078 & 0.77 \\
8 & -6.495 & -4.478 & 2.017 & -5.988 & -4.297 & 1.691 & 0.17 & 893 & 1.06 & 
6.12 & 0.14 & 1106 & 0.76 \\
9 & -6.495 & -4.478 & 2.017 & -5.688 & -4.597 & 1.091 & 0.12 & 902 & 1.05 & 
6.09 & 0.11 & 1127 & 0.75 \\
10& -6.495 & -4.478 & 2.017 & -5.388 & -4.897 & 0.491 & 0.09 & 907 & 1.05 & 
6.07 & 0.09 & 1141 & 0.75 \\
11& -6.495 & -4.478 & 2.017 & -5.088 & -5.197 & -0.109 & 0.09 & 909 & 1.04 & 
6.06 & 0.07 & 1149 & 0.75 \\
12& -6.495 & -4.478 & 2.017 & -5.134 & -5.134 & 0.000 & 0.09 & 909 & 1.04 & 
6.06 & 0.08 & 1149 & 0.75 \\
\hline
13& -5.486 & -5.486 & 0.000 & -5.134 & -5.134 & 0.000 & 0.04 & 873 & 1.13 & 
6.02 & 0.04 & 1101 & 0.81 \\
14& -5.134 & -5.134 & 0.000 & -5.134 & -5.134 & 0.000 & 0.04 & 837 & 1.22 & 
6.02 & 0.04 & 1059 & 0.87 \\
\hline
\end{tabular}
\end{table}

\begin{figure}[!t]
\begin{center}
\includegraphics[scale=0.8]{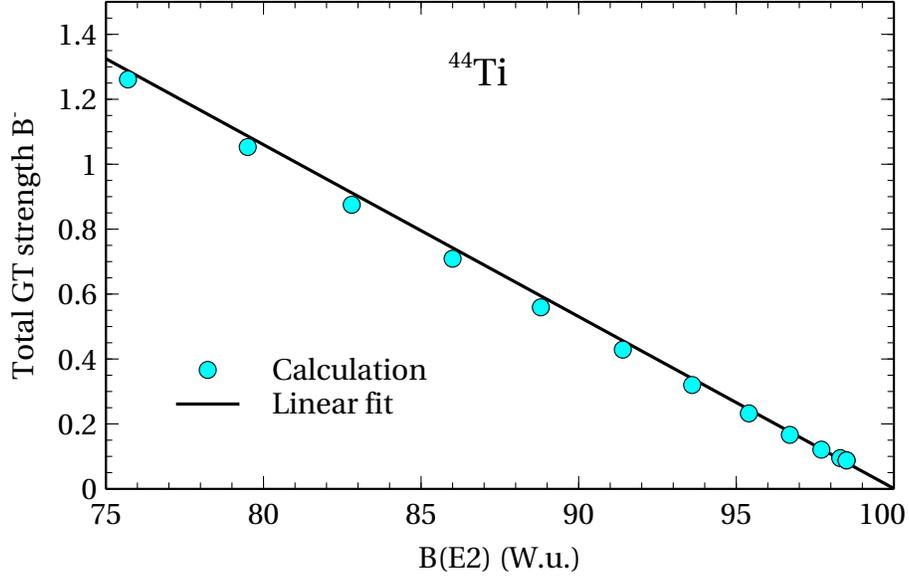}
\caption{The total GT strength from the ground state of $^{44}$Ti is linearly 
anticorrelated with the transition rate $B$(E2) (shown in Weisskopf units) from 
the ground state to the collective $2^{+}$ phonon state. \label{GT_BE2}}
\end{center}
\end{figure}

\begin{figure}[!t]
\begin{center}
\includegraphics[scale=0.8]{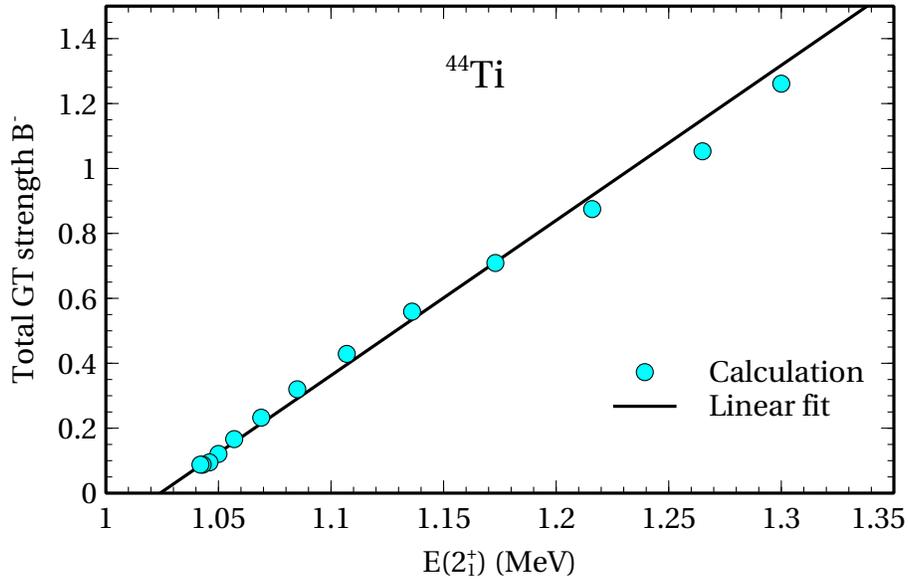}
\caption{Energy of the lowest quadrupole excitation in $^{44}$Ti is almost 
linearly anticorrelated with the total GT strength from the ground state when 
both are changed by the gradual elimination of the spin-orbit splitting. 
\label{GT_E2}}
\end{center}
\end{figure}

\begin{figure}[!t]
\begin{center}
\includegraphics[scale=0.8]{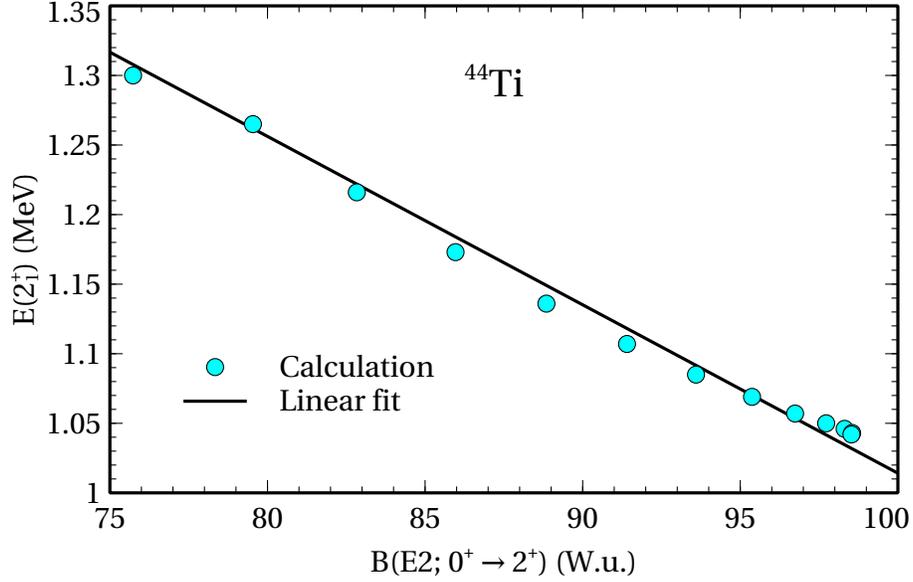}
\caption{Energy of the first collective quadrupole state in $^{44}$Ti is reduced 
while the corresponding quadrupole transition probability grows. 
\label{BE2_E2}}
\end{center}
\end{figure}

Table 1 and Fig.~\ref{GT_BE2} show the anticorrelation mentioned in the 
Introduction. In the isospin-symmetric nucleus
$^{44}$Ti, the total GT strength linearly falls to zero while $B$(E2) grows when the spin-orbit splitting
$\Delta\epsilon(f)$ is gradually reduced to zero. The weakening of the spin-orbit
coupling is harmful for the GT strength (in the limit of no such coupling, both GT strengths (\ref{3}) for $N=Z$ vanish,
see below). The energy of the quadrupole phonon $2^{+}$ state goes down, 
Fig.~\ref{GT_E2}, which is also reflected by the resulting increase of the 
quadrupole strength, Fig.~\ref{BE2_E2}. Qualitatively, we see a similar
evolution for $^{46}$Ti, where the sum rule (\ref{3}) gives 6.

\begin{figure}[!t]
\begin{center}
\includegraphics[scale=0.8]{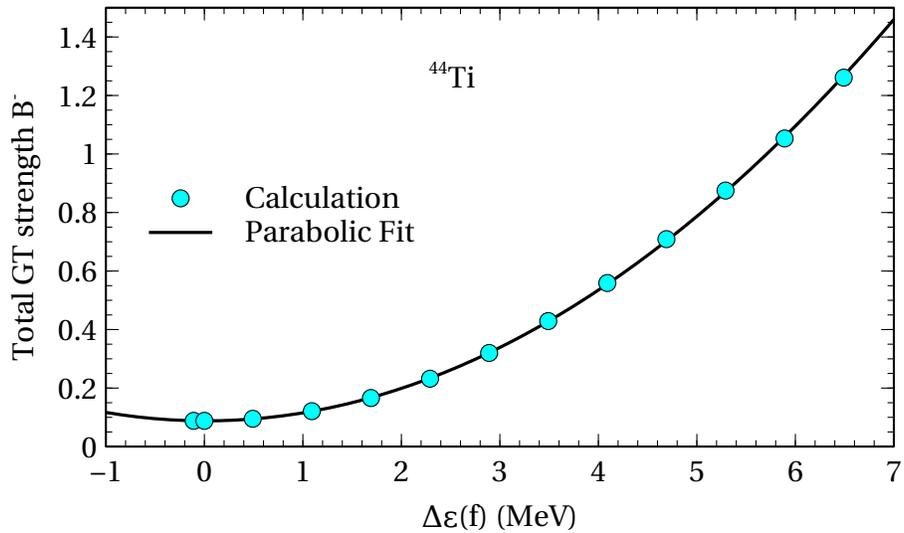}
\caption{The parabolic dependence of the GT strength on the spin-orbit splitting 
in $^{44}$Ti. \label{Ti44BGT_ELS}}
\end{center}
\end{figure}

\begin{figure}[!t]
\begin{center}
\includegraphics[scale=0.8]{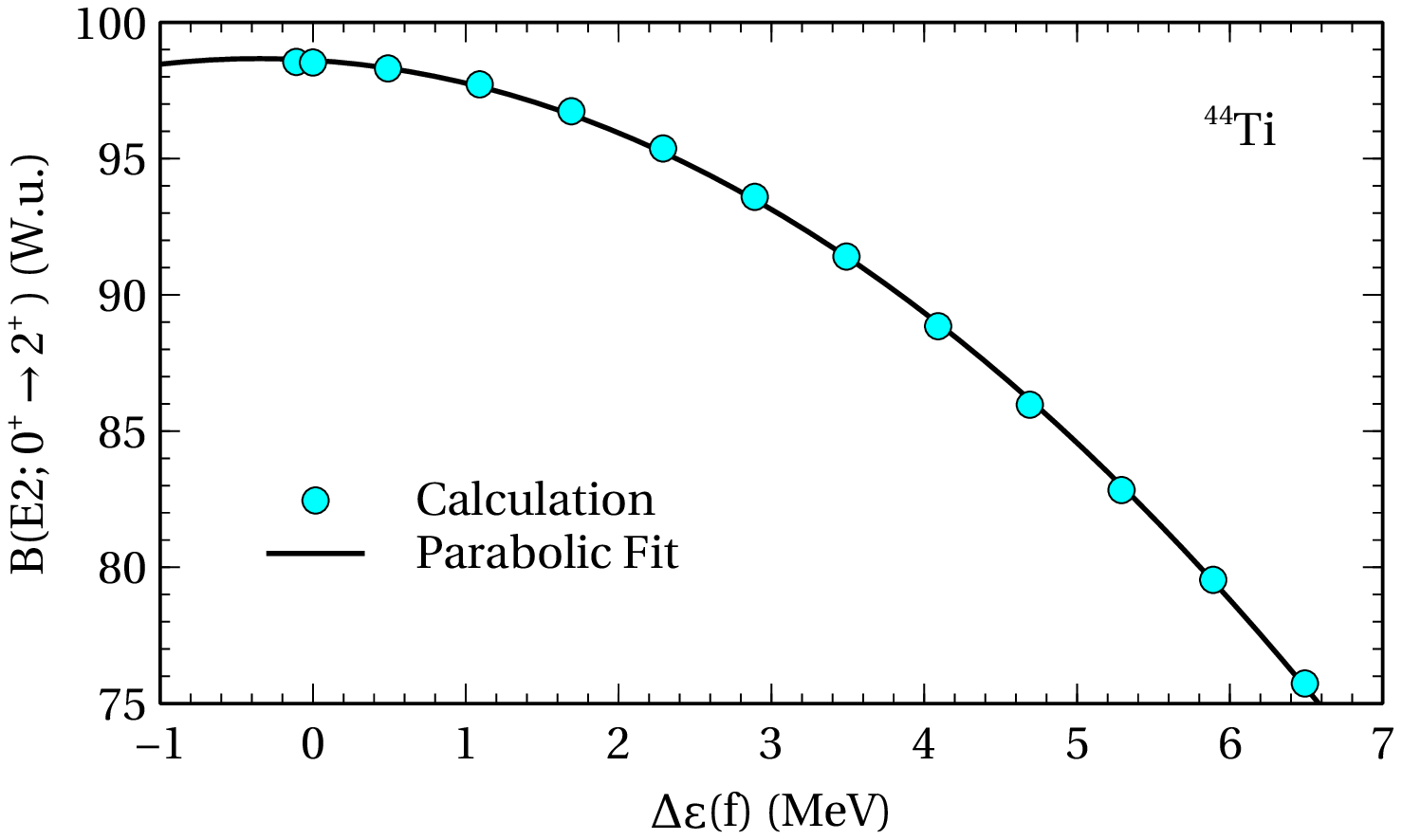}
\caption{The parabolic dependence of the quadrupole strength 
$B$(E2;$0^{+}\rightarrow 2^{+})$  on the spin-orbit splitting in $^{44}$Ti. 
\label{Ti44BE2_ELS}}
\end{center}
\end{figure}

\begin{figure}[!t]
\begin{center}
\includegraphics[scale=0.8]{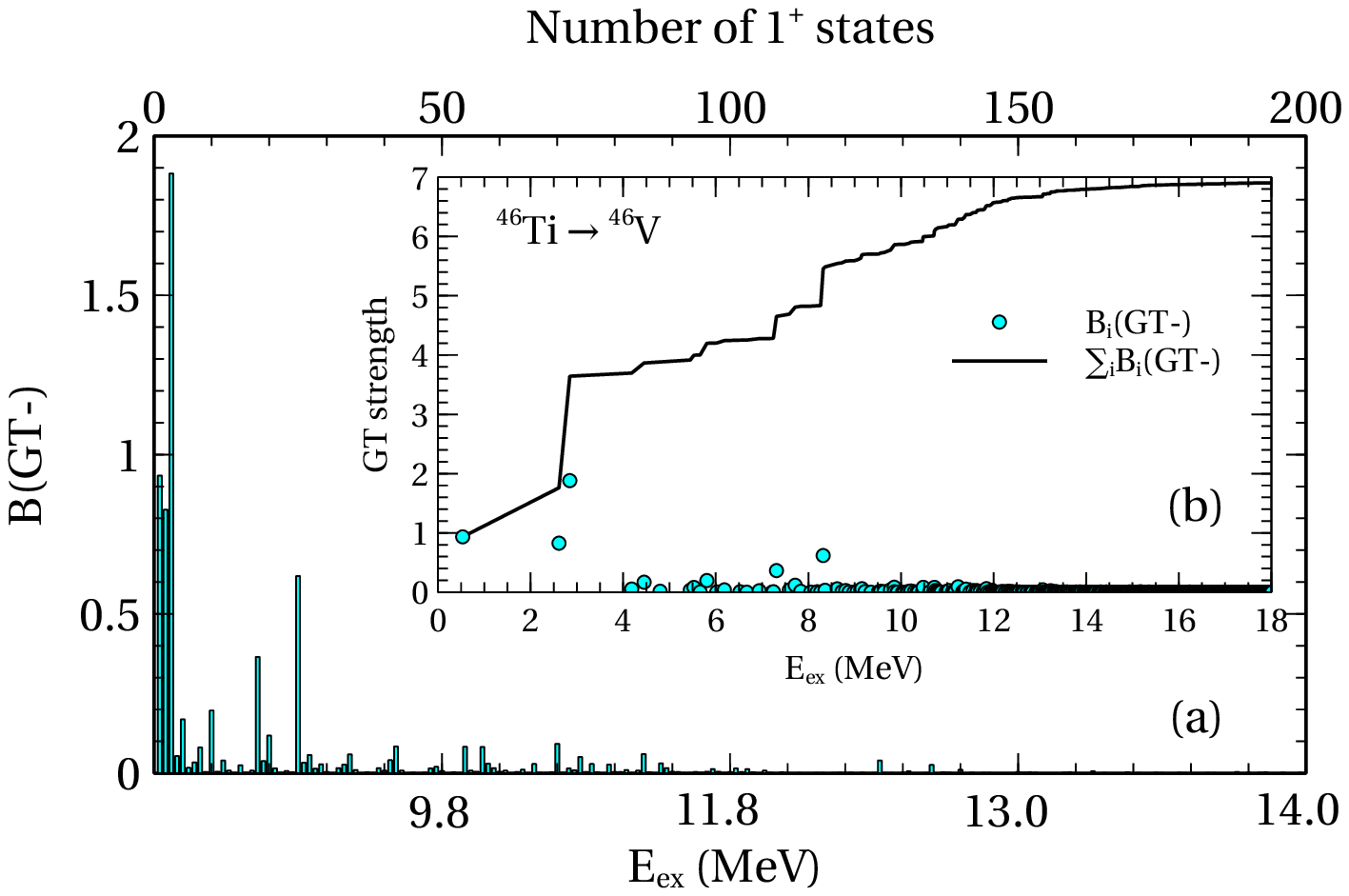}
\caption{(a) Distribution of the GT$^{-}$ strengths from the ground state of 
$^{46}$Ti along the excitation energy in the daughter state $^{46}$V. 
(b) Cumulative sum of the GT$^{-}$ strengths growing as a function of the 
excitation energy in the daughter nucleus $^{46}$V. \label{BGT_Eex_ni}}
\end{center}
\end{figure}

\begin{figure}[!t]
\begin{center}
\includegraphics[scale=0.8]{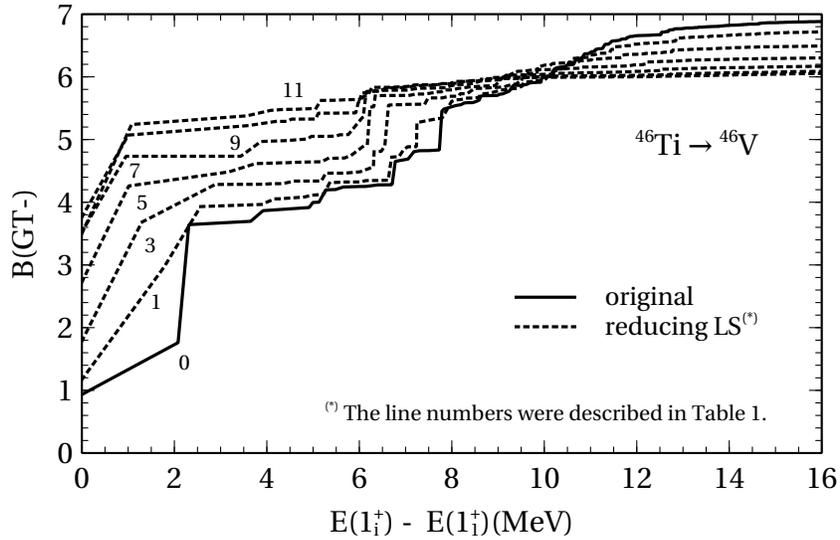}
\caption{Cumulative sum of the GT$^{-}$ strengths growing as a function of the 
excitation energy in the daughter nucleus $^{46}$V. The consecutive lines 
(labeled as the lines of Table 1) illustrate the accumulation process for 
several values of the spin-orbit splitting. \label{BGT_Eex_LS}}
\end{center}
\end{figure}

Figs.~\ref{Ti44BGT_ELS} and \ref{Ti44BE2_ELS} show that the changes of the 
summed GT strength and low-lying $B$(E2) transition probability 
as a function of the spin-orbit splitting $\Delta\epsilon(f)$ in $^{44}$Ti are almost exactly parabolic 
and opposite to each other. They do not depend on the sign of the spin-orbit 
coupling. Fig.~\ref{BGT_Eex_ni} illustrates the distribution 
of the GT$^{-}$ strength from the ground state of the $^{46}_{22}$Ti$_{24}$ nucleus in a function of 
the excitation energy in the daughter states of $^{46}$V. The same process of accumulating the total GT 
strength along the excitation energy of $^{46}$Ti is shown by 
Fig.~\ref{BGT_Eex_LS}; it works faster at small spin-orbit 
splitting, while in the realistic situation the accumulation of the total strength is going slower. 

\begin{figure}[!t]
\begin{center}
\includegraphics[scale=0.8]{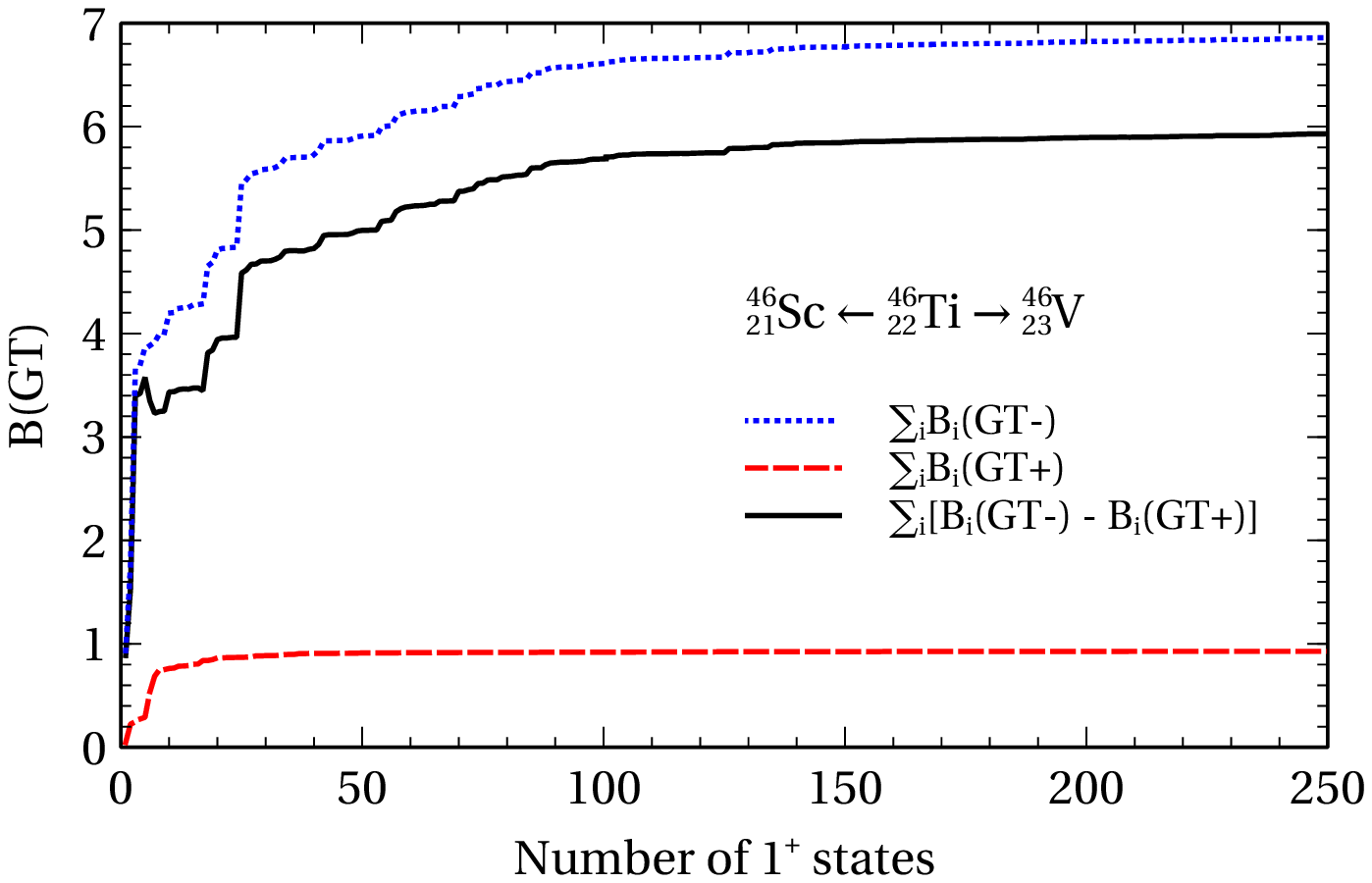}
\caption{Cumulative sum of the GT$^{-}$ and GT$^{+}$ strengths growing as a 
function of the number of $1^+$ states in the daughter nuclei $^{46}$V and 
$^{46}$Sc. \label{Ti46BGTmp}}
\end{center}
\end{figure}

This picture is practically universal, always the specific daughter states with a large GT strength do not 
give the full sum rule. Moving along the excitation energy of the daughter nucleus and collecting the GT strength 
from the mother state we can see the gradual filling of the total strength required by the GT sum rule.
Apart from few significant peaks in a cumulative sum, the convergence to the required value slowly proceeds 
through a large number of quite small increments. This can be seen in detail in 
Fig.~\ref{Ti46BGTmp}, where both the cumulative
strengths $B^{-}$ and $B^{+}$ for $^{46}$Ti are shown. The $B^{+}$ strength here 
is relatively small and quickly saturates, while the $B^{-}$ strength grows 
slowly until the difference sum rule (\ref{3}) is satisfied.

\section{Effect of spin-orbit splitting}

Here we comment on the spin-orbit coupling part of the mean field as an appropriate intermediary agent influencing both
 low-lying collective quadrupole vibrations and Gamow-Teller mode based on the spin excitation. Because of this coupling, the total orbital momentum $L$ of the excitation is not conserved, and one of the specific effects of spin-orbit coupling is the mixing of $L=0$ and $L=2$ excitations.

In agreement with findings of Ref. \cite{AZZB93}, in the limit of switched-off spin-orbit coupling,
the GT strength vanishes in $N=Z$ nuclei, $B^{-}=B^{+}=0$. This can be understood in terms of isospin
invariance and the $LS$ coupling scheme instead of the $jj$ coupling. Indeed, neutrons and protons occupy
here the same orbitals, so that the $n\leftrightarrow p$ transformations require the spin flip. This
changes the spin symmetry of the corresponding nucleon pair which could be compensated by the change of
orbital symmetry. However, if there is no coupling between orbital and spin momenta the process
turns out to be forbidden.

To illustrate this by the simplest example, consider the shell-model state of a valence $np$ pair that
should satisfy $(-)^{T+L+S}=-1$. For example take quantum numbers $L=0, \,S=0, \,T=1, \,T_{3}=0$ of the mother
state $|i\rangle, N=Z$,
\begin{equation}
|i\rangle=\,\frac{1}{\sqrt{2}}\,(p^{\dagger}_{1/2}n^{\dagger}_{-1/2}-p^{\dagger}_{-1/2}n^{\dagger}_{1/2})
|0\rangle,                                                       \label{4}
\end{equation}
where only spin projections of proton and neutron creation operators are indicated. The zero spin component
GT$_{0}^{-}$ {of the GT$^{-}$ operator acts as
\begin{equation}
({\rm GT})^{-}_{0}|i\rangle=\,\frac{1}{\sqrt{2}}\,[p^{\dagger}_{1/2}(s_{z}p^{\dagger}_{-1/2})-
p^{\dagger}_{-1/2}(s_{z}p^{\dagger}_{1/2})]|0\rangle=\,\frac{1}{2\sqrt{2}}[-p^{\dagger}_{1/2}p^{\dagger}_{-1/2}
-p^{\dagger}_{-1/2}p^{\dagger}_{1/2}]|0\rangle.                                    \label{5}
\end{equation}
Using the anticommutator of proton operators, we get zero. The ``down", GT$^{-}_{-}$, and ``up",  GT$^{-}_{+}$,   components of the GT operator do not act either:
\begin{equation}
({\rm GT})^{-}_{-}|i\rangle=-\,\frac{1}{\sqrt{2}}\,[p^{\dagger}_{-1/2}(s_{-}p^{\dagger}_{1/2})]=0,
                                                                   \label{6}
\end{equation}
Therefore in this case the GT strength vanishes, and it turns out the same for any even $L$. Now, for odd $L$ and $S=1$, we take $M_{L}=0,S_{z}=0$, and the result is the same. This negative mechanism works in a general case of $N=Z$ in the absence of spin-orbit coupling.

Table 1 and Figs. 1 and 2 show the monotonous growth of the GT strength for the $N=Z$ nucleus $^{44}_{22}$Ti$_{22}$ in the shell-model calculation for two valence $np$ pairs as a function of the increasing energy splitting between $f_{7/2}$
and $f_{5/2}$ orbitals. This splitting serves as a measure of the spin-orbit coupling strength.
At the same time, the $B$(E2) transition rate from the ground state naturally grows with the change of this splitting in the opposite direction due to the increasing softening of all simple transitions coupled into the collective mode.  \\

The typical spin-orbit term in the mean-field approximation can be written as
a sum of single-particle contributions,
\begin{equation}
H^{(ls)}=\sum_{a}h(r_{a})(\vec{\ell}\cdot{\bf s})_{a},                            \label{7}
\end{equation}
where the radial form-factor of spin-orbit coupling contains the radial derivative of the mean nuclear potential and  can be evaluated in average as
\begin{equation}
\bar{h}\approx -\,\frac{20}{A^{2/3}}\,{\rm MeV};                               \label{8}
\end{equation}
$|h|$ is slightly bigger in the shell-model description of the $pf$-shell nuclei used in our calculations.

The isoscalar quadrupole moment of the nucleus is taken as a sum over particles,
\begin{equation}
Q_{kl}=\sum_{a}(q_{kl})_{a}=\sum_{a}(3x_{k}x_{l}-r^{2}\delta_{kl})_{a}.       \label{9}
\end{equation}
The shift of the collective quadrupole excitation due to the spin-orbit splitting can be
estimated with the help of general arguments, for example using a simple model of factorizable
(in this case quadrupole-quadrupole) forces, $H_{Q}=-\kappa(Q\cdot Q)$.
As  discussed in textbooks, see for example \cite{ZV}, Section 18.1, in the case of an attractive
residual interaction, $\kappa>0$, the energy $\omega$ of a collective excitation is lower than the centroid of energies
$\bar{\epsilon}$ of independent (mean-field) excitations with the same quantum numbers,
$\omega \approx \bar{\epsilon}-\kappa {\cal N}\overline{q^{2}}$, where ${\cal N}$ is a characteristic
collectivity factor (a number of simple excitations coherently coupled to a collective mode) and
$\overline{q^{2}}$ their typical strength. A simplified model in Appendix B illustrates the main features
of the behavior of the collective frequency and transition rate seen in Table 1.

The GT strength from the ground state is, to a good approximation, a quadratic function of the spin-orbit
splitting. This is exactly what we should expect for transitions induced by a time-odd operator (magnetic dipole or GT).
As follows from the symmetry arguments (Ref. \cite{ZV}, Section 13.11), in such cases the matrix element
for the transition between orbitals $\lambda$ and $\lambda'$ is proportional to the combination
\begin{equation}
P^{(-)}_{\lambda\lambda'}= u_{\lambda}v_{\lambda'}- u_{\lambda'}v_{\lambda'},      \label{10}
\end{equation}
where the factors $u$ and $v$ describe the occupancies ($n_{\lambda}$ between zero and one)  of corresponding orbitals,
\begin{equation}
v_{\lambda}^{2}=n_{\lambda}, \quad u_{\lambda}^{2}=1-n_{\lambda}.         \label{11}
\end{equation}
For degenerate levels, the occupancies in equilibrium filling are equal, and the transition probability vanishes.
With spin-orbit splitting growing, the difference of occupancies grows quadratically with this splitting,
in agreement with what is given by the numerical calculation of 
Fig.~\ref{Ti44BGT_ELS}.

The spin-dependent contribution to the equation of motion for the quadrupole moment is found as
\begin{equation}
[H^{(ls)},Q_{kl}]=-3i\sum_{a}h(r_{a})\left([{\bf s}\times{\bf r}]_{k}x_{l}+[{\bf s}\times
{\bf r}]_{l}x_{k}\right)_{a}.                                              \label{12}
\end{equation}
Looking for the physical overlap of GT and quadrupole modes, we evaluate the double commutator
typical for the sum rules,
\begin{equation}
[V^{-}_{l},[H^{(ls)},Q_{kl}]]=3M^{-}_{k},        \label{13}
\end{equation}
where the sum over repeated Cartesian subscripts is assumed. The vector operators $M^{\pm}_{k}$ are
spin-quadrupole moments for the two opposite directions of the GT excitation,
\begin{equation}
M^{\pm}_{k}= \sum_{a}\tau^{\pm}_{a}h_{a}\Bigl(3({\bf s}\cdot{\bf r})x_{k}-r^{2}s_{k}\Bigr)_{a}.  \label{14}
\end{equation}
The physical effect of this dynamics is the appearance of the quadrupole component in the GT excitation
so that the part of the GT strength is now transferred to the $L=2$ channel.
In a crude estimate, the vectors (\ref{14}) are proportional to the original GT amplitudes.

For an estimate by order of magnitude we assume that the soft quadrupole mode with its direction
of slowly changing deformation generates on average the same directional character of the fast GT excitation,
so that $Q_{kl}\propto 3n_{k}n_{l}-\delta_{kl}$ and $V^{\pm}_{k}\propto v^{\pm}n_{k}$ in terms of the unit
vector ${\bf n}$. Then
\begin{equation}
[V^{+}_{k}V^{-}_{l},[H^{(ls)},Q_{kl}]]\;\Rightarrow\;6\bar{h}\bar{q}_{kl}n_{k}n_{l}\langle v^{+}v^{-}\rangle
=12\bar{h}\bar{q}\langle v^{+}v^{-}\rangle,                              \label{15}
\end{equation}
where the bar means the average over relevant single-particle transitions, and the matrix elements $q_{kl}$
were defined in eq. (\ref{9}). On the other hand, the sum rule following from the original equation of motion
with our auxiliary Hamiltonian, gives for the expectation value of the left hand side of eq. (\ref{15})
the estimate $4\Delta\omega Q\langle v^{+}v^{-}\rangle$. Here $\Delta\omega$ is the displacement of the collective
quadrupole excitation energy because of the spin-orbit splitting, and $Q$ is the phonon amplitude,
$Q={\cal N}\bar{q}$, where ${\cal N}$ is the factor of collectivity of the phonon excitation.
The comparison of two estimates gives
\begin{equation}
\Delta \omega\approx \,\frac{3\bar{h}}{Q/q}\,\approx -\frac{60}{A^{2/3}{\cal N}}\;{\rm MeV}. \label{16}
\end{equation}
This quantity is of the order 200-300 keV which is in agreement with 
Fig.~\ref{GT_E2}.

In this oversimplified approach, being mediated by the spin-orbit interaction, the centroid of the GT excitation
and the low-lying collective quadrupole excitation follow each other, in a qualitative agreement with exact 
results of shell-model computation. Fig.~\ref{BGT_Eex_ni} shows that the GT sum 
rule is getting fulfilled earlier in the process 
of gradual switching off the spin-orbit coupling when, as mentioned earlier, see eq. (\ref{16}) and Appendix B,
 the quadrupole frequency diminishes. The realistic spin-orbit interaction moves the GT final states up slowing 
the approach to the sum rule limit and making this process more fine-grained.

The whole interplay here can be considered as a result of the effective  interaction between quadrupole and GT and charge-exchage degrees of freedom, that, in the lowest order, can be written as
 $H_{{\rm eff}}\propto Q_{kl}V^{+}_{k}V^{-}_{l}$. This  is somewhat similar to the correlation between collective
octupole and quadrupole modes also described by the cubic anharmonic terms.
That correlation was predicted theoretically \cite{metlay95} and found experimentally \cite{mueller06} working practically
exactly for the chain of xenon isotopes. Later this effect was qualitatively observed in the data
for other isotope chains \cite{buchhorn09}. The same idea was useful in the theoretical search
\cite{ZVA08} for the enhancement of the nuclear Schiff moment, important in the problem of the
electric dipole moment, due to the combined action, and therefore correlation, of soft quadrupole
and octupole modes \cite{FZ03}.

\section{Conclusion}

We discussed some features of the nuclear GT processes which are not clearly formulated in the literature.
The phenomenon of anticorrelation between the GT strength and collectivity of the lowest quadrupole
excitation was studied numerically by exact shell-model calculations for the $fp$ orbital space and with the help of simple
clarifying models. The physics of this
phenomenon is based on Fermi statistics, isospin invariance and spin-orbit interaction.

In self-conjugate nuclei ($N=Z$) without spin-orbital splitting, the GT strengths in both directions would
vanish under exact isospin symmetry. This interrelation is illustrated by the shell-model calculations for
consecutive intermediate values of spin-orbit splitting. As follows from the general physics of low-lying
collective excitations, in the same process of eliminating spin-orbit splitting, the quadrupole frequency
goes down and the corresponding transition rate grows.

With restoration of the spin-orbit interaction, the GT strength centroid moves to higher energies with
increasing fragmentation. This process is anticorrelated with the enhancement of the collective quadrupole
mode. The limiting value of the universal GT sum rule is reached through growing fragmentation to many
weak transitions. The understanding of this process can again (see, for example, \cite{sakai})
raise the question of better evaluation of
experimental results on GT quenching with the detailed consideration of the significantly fragmented strength.

Another question that might reappear is the role of spin-orbit forces in mixing various values of the total
orbital momentum $L$ in GT processes and charge exchange reactions, including the isovector spin-monopole
giant resonance. In the presence of spin-orbit
coupling, the total orbital momentum $L$ of the nucleus is not conserved. With the spin-orbit coupling
as an intermediary, the GT pseudovector operator in the nuclear medium excites not only $L=0$ but also
$L=2$ states (these channels are interfering). The experimental treatment of charge-exchange reactions with the help of multipole decomposition typically
extracts from the angular distribution only the $L=0$ strength which does not reflect the total strength
excited by the GT operator inside the target nucleus. This question deserves
better attention from both experimental and theoretical viewpoints.

\section{Acknowledgements}

The authors are grateful to the Binational Science Foundation US-Israel for support that made possible
the completion of this work. V.Z. also acknowledges the support from the NSF grant PHY-1404442.
We thank B.A. Brown and R. Zegers for discussions.


\newpage

{\bf Appendix A. Operator algebra} \\
\\

The nine operators related to the ${\cal SU}$(4) group are
\begin{equation}
V^{\alpha}_{i}=\,\frac{1}{2}\,\sum_{a}(\sigma_{i})_{a}(\tau^{\alpha})_{a}.                   \label{A1}
\end{equation}
They commute (in Cartesian coordinates of vectors) according to
\begin{equation}
[V^{\alpha}_{i},V_{j}^{\beta}]=i\bigl(\epsilon_{ijk}\delta^{\alpha\beta}S_{k}
+\delta_{ij}\epsilon^{\alpha\beta\gamma}T^{\gamma}\bigr),             \label{A2}
\end{equation}
where ${\bf S}=\sum_{a}{\bf s}_{a}$ and ${\bf T}=\sum_{a}{\bf t}_{a}$ are the total spin (Latin subscripts)
and total isospin (Greek superscripts) operators, respectively. In particular (vector summation over $i$ in
the third equality),
\begin{equation}
[V^{+}_{i},V^{-}_{j}]=2i\epsilon_{ijk}S_{k}+2\delta_{ij}T^3, \quad [V_{i}^{\pm},V_{j}^{\pm}]=0, \quad
[V_{i}^{+},V_{i}^{-}]=3 (N-Z),                                                     \label{A3}
\end{equation}
in agreement with eq. (\ref{A3}). There are also simple ladder relations
\begin{equation}
[V^{\pm}_{i},T^{3}]=\mp V^{\pm}_{i}.                                         \label{A4}
\end{equation}

We note that this commutator of two vector operators, eqs. (\ref{A2}-\ref{A4}), contains only pseudovector
and scalar components with respect to spin coupling while the quadrupole component is absent. The squared
vector part is proportional to the GT intensity.\\

{\bf Appendix B. Simple model} \\
\\

Here we use an oversimplified but generic model to illustrate the conciliated behavior of the collective quadrupole 
frequency and corresponding transition probability under the change of spin-orbit splitting. Assume that we have 
two groups of degenerate single-particle levels (images of our $f_{7/2}$ and $f_{5/2}$ orbitals) with approximately 
the same single-particle matrix elements $q$ of the collective operator (a quadrupole moment in our problem). 
The interaction matrix elements $H'_{ij}$ are factorized as $\kappa q_{i}q_{j}$, where $\kappa<0$. The unperturbed Hamiltonian includes degenerate energies for those groups, $\epsilon_{1}=0$ and $\epsilon_{2}>0$, and pairing 
forces which create the energy gap $\Delta$ so that the characteristic excitation energies in an even system are 
$2\Delta$ and $2\sqrt{\Delta^{2}+\epsilon^{2}}$.

The secular equation for the collective energy $\omega$ contains the two groups of contributions:
\begin{equation}
1=\,\frac{S}{\omega-2\Delta}\,+\,\frac{S'}{\omega-2\sqrt{\Delta^{2}+\epsilon^{2}}},        \label{B1}
\end{equation}
where $S=\kappa \sum_{k}q_{k}^{2}$ and $S'=\kappa \sum_{k'}q_{k'}^{2}$ contain contributions of the first and the second groups of single-particle transitions, respectively. $S$ and $S'$ are quantities of the same order of magnitude and for simplicity we set $S=S'$. For typical numerical values of the upper line of Table 1, $\epsilon=6.5$ MeV, $\omega=1.3$ MeV and, in this region of the nuclear chart, $\Delta\approx$1.7 MeV, we extract $S\approx$ -1.8 MeV. Changing the level distance $\epsilon$ to zero we increase $\Delta$ and decrease the collective frequency $\omega$. Normalizing correctly the collective state \cite{ZV} we find  the collective transition probability at any point of this process,
\begin{equation}
B=\,\frac{4S}{\kappa}\,\frac{(\Delta+\sqrt{\Delta^{2}+\epsilon^{2}}-\omega)^{2}}
{(2\Delta-\omega)^{2}+(2\sqrt{\Delta^{2}+\epsilon^{2}}-\omega)^{2} }.                     \label{B2}
\end{equation}
The maximum of this probability is reached for degenerate levels, $\epsilon\rightarrow 0$, when
\begin{equation}
B_{{\rm max}}=\,\frac{2S}{\kappa}.                                                                         \label{B3}
\end{equation}
The ratio $B/B_{{\rm max}}$ for the upper line of Table 1 is predicted by eq. (\ref{B2}) to be 0.67 which agrees with the numerical results in this table.

\end{document}